\documentclass{elsart}
\usepackage{graphicx,natbib,amssymb,rotating}
\journal{J. of Advances in Space Research}
\begin{document}
\begin{frontmatter}
\title{Specific interplanetary conditions for CIR-, Sheath-, and ICME-induced 
       geomagnetic storms obtained by double superposed epoch analysis}
\author[IKI]{Yu.I. Yermolaev\corauthref{cor}},
\corauth[cor]{Corresponding author. 
              Tel.: +7 495 3331388, Fax: +7 495 3331248}
\ead{yermol@iki.rssi.ru}
\ead[url]{www.iki.rssi.ru/people/yyermol$\_$inf.html}
\author[IKI]{N.S.~Nikolaeva
}
\ead{nnikolae@iki.rssi.ru},
\author[IKI]{I.G.~Lodkina
}
\ead{lodkina@iki.rssi.ru},
\author[IKI]{M.Yu.~Yermolaev
}
\ead{yermol@iki.rssi.ru}
\address[IKI]{Space Plasma Physics Department, Space Research Institute (IKI), 
Russian Academy of Sciences, Profsoyuznaya 84/32, Moscow 117997, Russia}

\begin{abstract} 

A comparison of specific interplanetary conditions for 798 magnetic storms 
with Dst $<$ -50 nT for the period 1976-2000 was made on the basis 
of the OMNI archive data. We categorized various large-scale types of solar wind
as interplanetary drivers of storms: corotating interaction region (CIR), Sheath, 
interplanetary CME (ICME) including magnetic cloud (MC) and Ejecta, separately MC 
and Ejecta, and "Indeterminate" type. The data processing was carried out by 
the method of 
double superposed epoch analysis which uses two reference times (onset of storm 
and the minimum Dst index) and makes a re-scaling of main phase of storm in a such way 
that 
all storms have equal durations of the main phase in the 
new time reference frame. This method reproduced some well-known results and allowed 
us to obtain some new results. Specifically, 
obtained results demonstrate high importance of 
Sheath in generation of magnetic storms as well as a significant differences in 
properties of MC and Ejecta and in their geoeffectiveness.

\end{abstract}

\end{frontmatter}

\section{Introduction} 

One of the important aims of Space Weather is investigation of interplanetary drivers of magnetic storms. It has been well known for a long time that the most important parameter leading to geomagnetic disturbances and, in partially, to  magnetic storm generation is negative (southward) Bz component of interplanetary magnetic field (IMF) 
\citep{FairfieldCahill1966,RostokerFalthammar1967,Russell1974,Burton1975,Akasofu1981}.
Because IMF lies in ecliptic plane under steady interplanetary conditions and substantial $Bz<0$ is observed only in disturbed types of solar wind (SW), 
it was found in many investigations that interplanetary coronal mass ejections (ICME) and corotating interaction regions (CIR) are the most important drivers of magnetic disturbances on the Earth. Therefore, it is natural to categorize these solar wind drivers during a study of magnetic storm generation (see reviews and recent papers, for instance, 
\citep{TsurutaniGonzalez1997,Gonzalez1999,YermolaevYermolaev2002,HuttunenKoskinen2004,Yermolaev2006,BorovskyDenton2006,Denton2006,Huttunen2006,Yermolaevetal2007a,Yermolaevetal2007b,Yermolaevetal2007c,Pulkkinen2007a,Pulkkinen2007b}
and references therein).  Nevertheless, the question what interplanetary conditions results in specific magnetic storm features is still open and it does not allow us to precisely predict reaction of the magnetosphere. 
The progress in solving this problem may be connected  with development of methodical approaches. 

Usage of the "peak-to-peak" method when minimum of IMF $Bz<0$ (or electric field $Ey = Vx * Bz$) was compared with extreme values of the Dst and Kp indexes did not allow one to find significant differences between these interplanetary and magnetospheric parameters for magnetic storms generated by different drivers (see, for instance, 
\citep{Yermolaevetal2007c}
and references therein). Analysis of time evolution in interplanetary parameters using the superposed epoch analysis (SEA) method was more informative and showed  some characteristic conditions during storms and several differences in interplanetary parameters for different drivers (see Table 1). Although the CIR, ICME and Sheath (compressed region before ICME) as important drivers of magnetic storm have been discussed in the literature for a long time  
\citep{Gosling1991,TsurutaniGonzalez1997,Vieira2004,HuttunenKoskinen2004,Yermolaev2005,YermolaevYermolaev2006,
Alves2006,BorovskyDenton2006},  
still some authors 
did not separate large-scale types of solar wind for superposed epoch analysis 
\citep{Denton2005,Zhang2006,Liemohn2008,Ilie2008}
or Sheath and body of ICME 
\citep{Denton2006,BorovskyDenton2006}. 
Only recent papers analyzed separately CIR, Sheath and body of ICME  
\citep{HuttunenKoskinen2004,Yermolaev2006,Huttunen2006,Yermolaevetal2007a,Yermolaevetal2007b,Yermolaevetal2007c,Pulkkinen2007a}.
The choice of zero (reference) time for SEA is important and substantially influences on 
results 
\citep{Yermolaev2006,Yermolaevetal2007a,Ilie2008}.
In the most part of previous papers the authors used the peak of Dst index as a zero time for SEA
\citep{Maltsevetal1996,LoeweProlss1997,Zhang2006}. 
This choice is convenient for studying the end of main phase and beginning of recovery phase of storms because 
duration of the main phase lasts from 2 to 15 hours  
\citep{Vichare2005,GonzalezEcher2005,Yermolaev2006,Yermolaevetal2007b}, 
inside of an interval with duration of several hours 
the parameters measured before and after onset are averaged simultaneously 
and specific conditions resulting in storm onset cannot be studied. 
The onset time as zero time of SEA allows one to investigate interplanetary sources and initial part of storms 
\citep{Yermolaev2006,Yermolaevetal2007a,Yermolaevetal2007b,Yermolaevetal2007c,Pulkkinen2007a}. 
For example, this analysis showed that storms initiated by Sheath have sharper and shorter main phase 
than storms initiated by another interplanetary drivers   
\citep{Yermolaevetal2007b,Pulkkinen2007a}.

In this paper we study interplanetary conditions resulting in magnetic storms on the basis of the OMNI database  
during 1976-2000 and categorize 6 types of solar wind: (1) CIR, (2) ICME, (3,4) two types of ICME -– Magnetic cloud (MC) and Ejecta, (5) Sheath before MC 
and Ejecta, and (6) "Indeterminate" type (OMNI does not contain sufficient information for identification of the type). 
We use "double" (with two reference times) ESA method, that is, we re-scale duration of main phase of all storms in a such manner that, respectively, onsets and minima of Dst index for all storms coincide and study interplanetary conditions leading to start and end of main phase of magnetic storms induced by these 6 interplanetary drivers.  

\section{Method} 

The basis of our investigation is 1-h interplanetary and magnetospheric data of OMNI database 
\citep{KingPapitashvili2004}.
We made our own data archive including OMNI data and calculated (using OMNI data) additional parameters. Using level criteria for key parameters of SW and IMF (velocity, temperature, density, ratio of thermal to magnetic pressure, magnitude and orientation of magnetic field etc.) we defined corresponding large-scale types of SW and possible error of this identification for every 1-h point of the archive during 1976-2000 
\citep{Yermolaevetal2009a}.
Our identification of SW types is based on 
methods 
similar to ones described in 
many papers (see reviews by 
\cite{Wimmer2006} and
\cite{Tsurutani2006} 
and references therein) and 
basically agrees with results of other authors but in contrast with other similar studies, 
we used general set of level criteria for all SW types and made identification for each 1-hour point.

During 1976-2009 there were 798 magnetic storms with $Dst < -50$ nT. There were data gaps in several parameters of OMNI database for 334 (42\%) from them (these storms are denoted as "Indeterminate" (IND)) and interplanetary drivers were found for 464 storms
\citep{Yermolaevetal2009b}. 
Magnetic storm is considered to be connected with specific SW type if its onset is observed in 2 hours after beginning and during this SW type (2 hours interval is the average delay between appearance of southward IMF and reaction of magnetosphere 
\citep{GonzalezEcher2005,Yermolaevetal2007a,Yermolaevetal2007b}.
THe statistics of magnetic storm distribution over different types of SW is following: 
145 storms were induced by CIR, 62 - MC, 161 - Ejecta, 96 -Sheath (12 before MC and 84 before Ejecta) 
\citep{Yermolaevetal2009b}.
In our previous papers 
\citep{Yermolaev2006,Yermolaevetal2007a,Yermolaevetal2007b,Yermolaevetal2007c}
we considered magnetic storms with $Dst < -60$ nT and statistics of storms was a bit different 
(total number of storms is 623).
 
As has been indicated above the duration of main phase of magnetic storms varies in a wide range of 2-15 hours 
\citep{Vichare2005,GonzalezEcher2005,Yermolaev2006,Yermolaevetal2007b},
average duration in the interval 1976-2000 is $7 \pm 4$ hours 
\citep{Yermolaev2006,Yermolaevetal2007b}.
We used SEA with 2 reference time instants and located all onsets at "0" time of epoch and all Dst minima at "6" time. The times before onset 
($t \le $ "0" time) and after Dst minimum 
($t \ge $ "6" time) are real, but the time between onset and Dst minimum was re-scaled (proportionally increased/decreased). 
After this transformation all storms have equal durations of main phase in the 
new time reference frame. 
This interval of main phase has 5 sub-intervals. 
The time in the "0" -- "6" interval 
for 2/3 storms was changed not more than by 1/3 of its duration. This "double" SEA method allows us to simultaneously study interplanetary conditions resulting in beginning and end of magnetic storms as well as dynamics (temporal variations) of parameters for storms with different durations. 

Taking into account that typical durations of large-scale SW types are significantly shorter than the full duration of magnetic storms (see, for example, 
\cite{Crookeretal1999};  
\cite{GoslingPizzo1999};  
\cite{Lynchetal2003};  
\cite{Leppingetal2005}) and in the average they are $9 \pm 4$, $28 \pm 12$ and $20 \pm 8$ hours for Sheath, ICME and CIR, respectively 
\citep{Yermolaevetal2007a},
we restricted durations of curves calculated using double SEA by time from -12 up to +24 hours. It is important that statistics decreases from main phase of storms to edges of (-12, +24) interval (especially for Sheaths) and errors may increase at the edges of interval (Standard deviations for different parameters and different interplanetary drivers are similar to data presented by 
\citep{Yermolaevetal2007a}). 
In 
cases discussed below differences between the curves in figures are 
mathematically significant although sometimes they are 
less than corresponding standard deviations. 
In some another cases it is necessary 
to consider these differences as a tendency (hypothesis) rather than a proven 
physical fact. The further investigations are required to reduce this uncertainty. 

\section{Results}     

Figures 1 and 2 present time variation of several interplanetary and magnetospheric parameters for 798 magnetic 
storms with Dst $<$ -50 nT during 1976-2000, which were 
obtained by the double SEA method with 2 reference epoch zero times:
Dst storm onset and Dst minimum (dashed lines which cross "0" and "6" in the time axis of figures, respectively) 
and for 6 interplanetary drivers: (1) MC, (2) Ejecta, (3) sum of MC and Ejecta, 
(4) Sheath (sum of Sheaths before MC and Ejecta), (5) CIR, and (6)  IND ("Indeterminate") type of SW. 
Curves for different types of solar wind are presented by 
different symbols/color.
Under figures designations and number of events for each SW type are specified, the number of points in a separate bin of curves in figures can be less than the specified number of events, especially at the interval edges. 
Figure 1 shows: (Left column) $V$ - velocity, $T$ - proton temperature, $T/Texp$ - 
ratio of measured proton temperature to calculated temperature $Texp$ using 
average dependence of temperature on velocity $V$
\citep{LopezFreeman1986}, 
Dst index, (Right) $n$ - density, $Pt$ - thermal pressure, 
$Pdyn$ - dynamic pressure, Kp index. 
Figure 2 presents other parameters: (Left column) $Ey$ - electric field, $Bz$ -  GSM southward components of IMF, 
$Dst^*$ - density corrected Dst index 
\citep{Burton1975}, 
$AE$ index, (Right) $By, Bx, B$ - GSM components and magnitude of IMF, $\beta$ - ratio of proton thermal to magnetic  
pressure.  

For all interplanetary drivers of magnetic storms the onset starts in 1-2 hours after southward turning of IMF ($Bz <$ 0)  and the main phase stops in 1-2 after disappearance of the southward IMF component. In new time scale in region of "0" - "6" there is no significant difference between behaviour of Dst (Dst*) index for different drivers. Nevertheless, it is possible to indicate a slight tendency that most sharp decrease of indices is observed for Sheath and MC while the largest values of southward component of IMF and electric field are in MC. The highest values of Kp and AE indices are generated by Sheath and MC. Slope of Dst (Dst*) index and values of Kp and AE indices for Ejecta and MC+Ejecta are less than for MC, and this fact is one of the most important reasons why we did not consider all ICME together and made selection of two subtypes of ICME: MC and Ejecta. The highest value of velocity $V$ is observed in Sheath (difference relative to another curves is 70-100 km/s), temperature $T$ and $T/Texp$ (2-3 times) in Sheath and CIR, density $n$ (1.5-2.0 times) in Sheath and CIR, thermal pressure $Pt$ (5-7 times) in Sheath, dynamic pressure $Pdyn$ (1.5-2.0 times) in Sheath and CIR, $\beta$ (1.5-2.0 times) in Sheath and CIR, IMF magnitude $B$ (2-5 nT) in Sheath, MC and CIR. There is no systematic difference in $Bx$ and $By$ components of IMF for different SW types in the region of "0" -- "6" time. 

To study behaviour of interplanetary parameters separately during moderate and strong storms we divide data presented in Figs.1 and 2 in two groups with -100 $< Dst <$ -50 nT (see Figs. 3 and 4) and $Dst <$ -100 nT (Figs. 5 and 6). This selection decreased data statistics in the new figures, especially for Figs. 5 and 6 (see number of events under figures) when number of points in a separate bin of curves at interval edges can be less than 10. Taking into account accuracy of our estimations, it is possible to tell that all specified above tendencies and features of relative behaviour of SW and IMF parameters and magnetospheric indices for different interplanetary drivers 
are the same for both storm sizes.

One of important problems of connection between interplanetary conditions and magnetospheric processes is dependence of magnetospheric activity on temporal evolution of SW and IMF parameters including $Bz$ and $Ey$. We found a consistency between time evolution of cause ($Bz$ and $Ey$) and time evolution of effect ($Dst, Dst^*, Kp$ and $AE$ indices) for the time interval of "0" - "6" using 3 procedures (see Figs.7-9): 
(1. Left column in figures where abscissa is designated as $Ey$ and $Bz$) simultaneous measurements, for example, dependence of $Dst(t^i).vs.Ey(t^i), i=0,...,6,$ 
(2. Second column, abscissa is $Ey(t-1)$ and $Bz(t-1)$) 1-h displaced measurements, for example, 
$Dst(t^i).vs.Ey(t^{i-1}), i=0,...,6,$ and 
(3. Right column, abscissa is $Ey(\sum)$ and $Bz(\sum)$) dependence of indices on integral value of sources, for example, $Dst^i.vs.Ey(\sum)^{i}= \int_{0}^{t^{i}} Ey(\tau) d\tau = \sum_0^i Ey^{k}, i=0,...,6; k=0,...,i$. It is important to remind that all storms have equal durations in interval of "0" --"6".  

The left column of Fig.7 shows that there is no any monotonic relation between interplanetary conditions and magnetospheric indices. If we take into account the 1-h delay between sources and effects (the second column), dependencies become more monotonic, i.e. there is delay between cause and effect. 
The right column of figure demonstrates that discussed processes have a "memory" and all dependencies are monotonic and almost linear. It is important to note that the most effective process of Kp and AE indices generation acts during Sheath, CIR and IND, the least effective process of Dst and Dst* generation - during MC. Additional selection of Fig.7 data on storm size with -100 $< Dst <$ -50 nT  and $Dst <$ -100 nT (see Figs. 8 and 9, respectively) shows that main properties specified in Fig.7 remain in the new figures.   

\section{Discussion and Conclusions}

We performed an analysis of interplanetary conditions for 798 magnetic storms 
with Dst $<$ -50 nT for the period 1976-2000 on the basis of the OMNI archive data. 
Our analysis has two special features. 
(1) Taking into account importance of epoch time selection we used the method of 
the double superposed epoch analysis
including simultaneously two reference times: the onset of storm and the minimum Dst index. 
(2) Taking into account different reaction of magnetosphere on 
various interplanetary disturbances we categorized large-scale types 
of solar wind as interplanetary drivers of storms: CIR, Sheath, MC, Ejecta, ICME 
(e.g. both MC and Ejecta) and "Indeterminate" type. 
This methodical approach showed the following.  

First of all we would like to note that our method reproduced some well-known results and, 
in particularly, showed that independently on types of interplanetary drivers, the onset 
of magnetic storms begins in 1-2 hours after southward turn of IMF and recovery phase 
of storms begins in 1-2 hours after disappearance of this component of IMF. This is a good 
verification of used method. At the same time the method allowed us to obtain some new results.  

Various types of interplanetary drivers of magnetic storms have significantly different parameters.
Particularly, Sheath and CIR (with respect to ICME) have higher density, dynamic and thermal pressures, temperature, $\beta$-parameter 
as well as higher variance of the same parameters and magnitude and components 
of IMF. 
These differences can be significant, for example, dynamic pressure is larger by a factor of 
1.5 - 2. 
Our analysis confirms high importance of Sheath in generation of magnetic storms
\citep{HuttunenKoskinen2004,Yermolaev2006,Huttunen2006,Yermolaevetal2007a,Yermolaevetal2007b,Yermolaevetal2007c,Pulkkinen2007a} 
and indicates specific geoeffective conditions in Sheath. 
It is important to note that there are serious differences in parameters of MC and Ejecta (including their geoeffectiveness), while differences in Sheath before MC and Sheath before Ejecta are not significant. This fact should be taken  into account during analysis. 

Using double SEA method (transformation of main phases of storms to equal duration) allowed us to compare storms with different durations of main phase. In the re-scaled temporal reference frame the Dst.vs.time (Dst*.vs.time) dependencies for storms induced by different types of interplanetary drivers are close to each other and have approximately linear shapes in time interval of "0" –- "6". Comparison of Dst.vs.time and Dst*.vs.time dependencies for storms induced by Sheath showed that dynamic pressure $Pdyn$ results in parallel displacement of Dst(Dst*).vs.time dependencies but does not change slope of these dependencies. Though our transformation should mask the shorter main phase for Sheath-induced storms
\citep{Yermolaevetal2007b,Pulkkinen2007a},
our results speak in favor that the Dst.vs.time (Dst*.vs.time) dependencies for storms induced by Sheath are  more abrupt. The Kp.vs.time and AE.vs.time dependencies are nonlinear and higher for Sheath- than for CIR- and ICME-induced storms. It means that these indices are generated by other mechanisms than Dst and Dst* indices.   

Dependencies of Dst (or Dst*) on the integral of Bz over time are almost linear and parallel for different types of drivers. 
This fact can be considered as an indication that time evolution of main phase of storms depends not only on current
values of $Bz$ and $Ey$ but also on their prehistory. 
The differences between these lines are relatively small ($\Delta Dst < 20$ nT), nevertheless there is tendency that the MC curve lies higher (i.e. at equal values of integral of Bz the storm is smaller). Dependencies on integral of Ey are similar. 
 Dependencies of Kp (and AE) on integral of Bz (and Ey) over time are nonlinear and nonparallel. The differences between these lines are relatively small ($\Delta Kp < 1$ and $\Delta AE < 50$ nT), nevertheless there is tendency that the Sheath and CIR curves lie higher than for another drivers. 

\section{Acknowledgments} 
The authors thank the OMNI database team for available data on the 
interplanetary medium and magnetospheric indices.
      The work was in part supported by RFBR, grants 04-02-16131 and 07-02-00042 and by 
Russian Academy of Sciences, programs
"Plasma Processes in the Solar System" and "Solar Activity and
Physical Processes in the Sun-–Earth System".  

\newpage

\begin{table}
\caption{List of results on interplanetary conditions resulting in magnetic storms 
obtained by superposed epoch analysis}  
\label{table:1}      
\centering           
{\scriptsize 
\begin{tabular}{l l l l l l }   
\hline 
N  & Number(Years) & Zero time & Selection & SW and IMF & Reference \\ 
\hline
1 & 538(1963-1991)& onset & No & B,Bx,By,Bz,V,T,n,Pdyn& 
\cite{Tayloretal1994}\\ 
2 & 120(1979-1984)&  min Dst & No & Bz, n, V & 
\cite{Maltsevetal1996}\\
3 & 150(1963-1987)& turning Bz& No& Bz,Pdyn& 
\cite{Davisetal1997} \\
4 & 305(1983-1991)& onset & No & Bz, Pdyn&  
\cite{YokogamaKamide1997} \\
5 & 1085(1957-1993)& min Dst & Dst & Bz, Pdyn &  
\cite{LoeweProlss1997} \\ 
6 & 130(1966-2000)& onset & No &B,Bx,By,Bz,$|Bx|,|By|$,& \\
  &               &       &    &$|Bz|$,V,n,Pdyn & \cite{LyatskyTan2003} \\   
7 & 623(1976-2000)& onset and min Dst& SW types$^a$& B,Bx,By,Bz,V,T,n,Pdyn, & \\
  &               &        & & nkT, $\beta$, T/Texp& 
\cite{Yermolaev2005,Yermolaev2006,Yermolaevetal2007a,Yermolaevetal2007b}\\
8 & 78(1996-2004) & min Dst& SW types$^b$& B,Bz,dB$/$B,V,T,n,& 
\cite{MiyoshiKataoka2005} \\ 
9 & 549(1974-2002)& min Dst&  Yes$^c$& B,Bx,By,Bz,$|Bx|,|By|$,& \\   
  &               &        & & $|Bz|$,Bs,VBs,V,n,T,Pdyn & \cite{Zhang2006} \\  
10& 623(1976-2000)& onset & SW types$^a$& $\sigma B$, $\sigma V$, $\sigma T$, $\sigma n$& 
\cite{Yermolaevetal2007c}\\
11& 28(1997–2002)& onset and min Dst& SW types$^d$& Bz, Pdyn, V, Ey & 
\cite{Pulkkinen2007a}\\
12& 10(2004)& onset& No & Bx,By,Bz,B,$\epsilon$,V,n,Pdyn, & \\
  &         &      &    & $M_A$,Ey &   
\cite{Pulkkinen2007b}\\
13& 29(1999-2002)& onset,main phase, & No& Bz,Pdyn& \\
  &              & min Dst           &   &        &
\cite{Ilie2008}\\
\hline
\multicolumn{6}{l}{$^a$ - (1) CIR, (2) Sheath and (3) MC; 
$^b$ - (1) CIR and (2) MC (Sheath + MC body); }\\ 
\multicolumn{6}{l}{$^c$ - (1) moderate storm at solar minimum, 
(2) moderate storm at solar maximum, (3) strong storm }\\
\multicolumn{6}{l}{ at solar minimum, and (4) 
strong storm at solar maximum; $^d$ - (1) MC and (2) Sheath.} \\
\hline
\end{tabular} 
}
\end{table}


\begin{figure}
\includegraphics[width=0.8\textwidth]{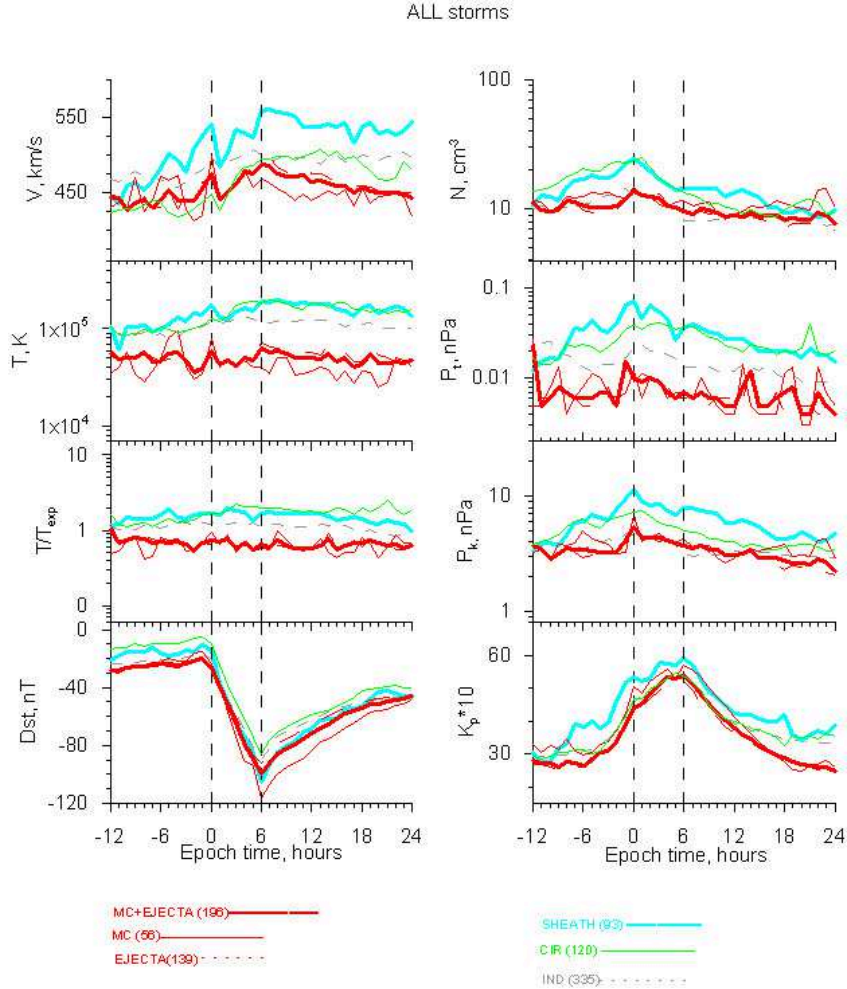}
\caption{Behavior of interplanetary parameters and magnetospheric Dst and Kp indices for magnetic storms with Dst $<$ -50 nT generated by different interplanetary drivers: (1) all ICME (MC+Ejecta), (2) MC, (3) Ejecta, (4) Sheath, (5) CIR, and (6) "Indeterminate" (see designations in bottom of figure)
 during 1976-2000 on the basis of OMNI database obtained by double  superposed epoch analysis method with two reference times: onset ("0" time, 1st dashed line) and minimum Dst index ("6" time, 2nd dashed line). Presented parameters: (Left column) $V$ - velocity, $T$ - proton temperature, $T/Texp$ - 
ratio of measured proton temperature to calculated temperature $Texp$ using 
average dependence of temperature on velocity $V$
\citep{LopezFreeman1986}, 
Dst index, (Right) $n$ - density, $Pt$ - thermal pressure, 
$Pdyn$ - dynamic pressure, Kp index.}
\end{figure}

\begin{figure}
\includegraphics[width=0.8\textwidth]{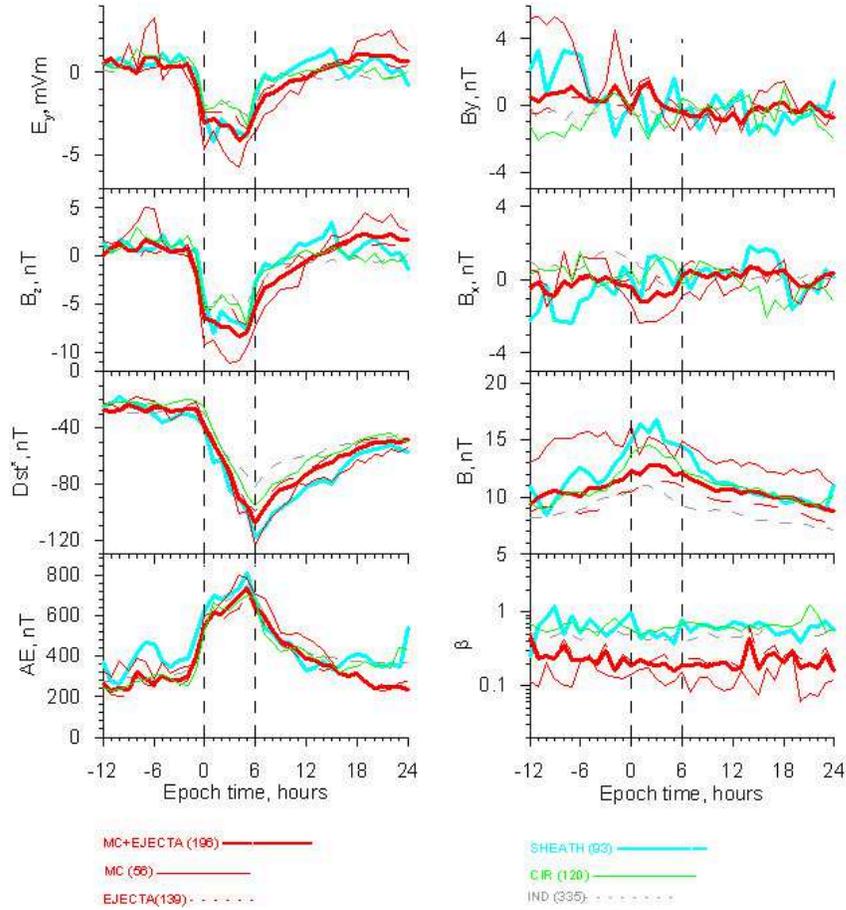}
\caption{The same as in Fig.1 for parameters: (Left column) $Ey$ - electric field, $Bz$ -  GSM southward components of IMF, $Dst^*$ - density corrected Dst index 
\citep{Burton1975}, 
$AE$ index, (Right) $By, Bx, B$ - GSM components and magnitude of IMF, $\beta$ - ratio of thermal to magnetic  
pressure.}
\end{figure}

\begin{figure}
\includegraphics[width=0.8\textwidth]{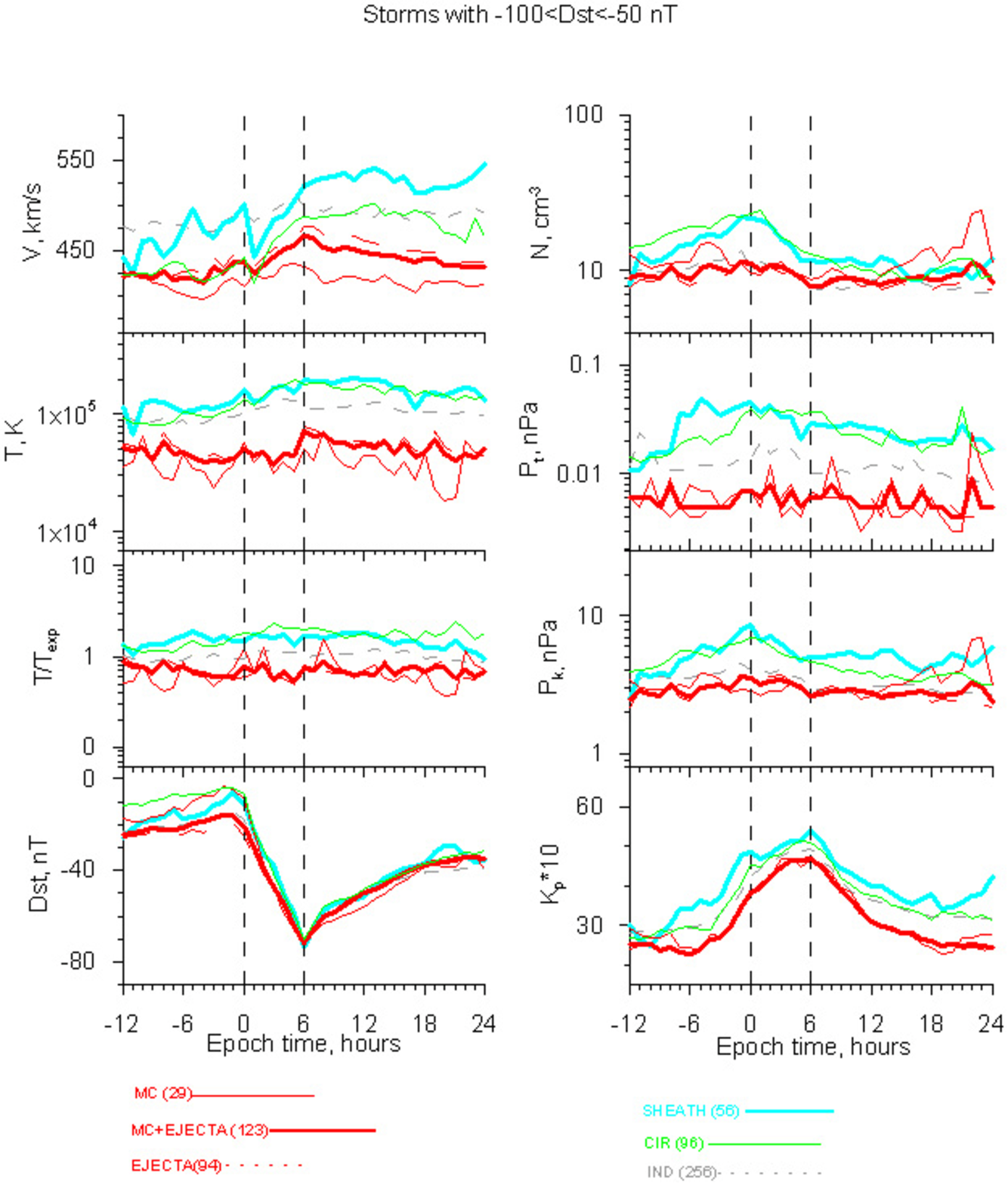}
\caption{The same as in Fig.1 for moderate storms with -100 $< Dst <$ -50 nT.}
\end{figure}

\begin{figure}
\includegraphics[width=0.8\textwidth]{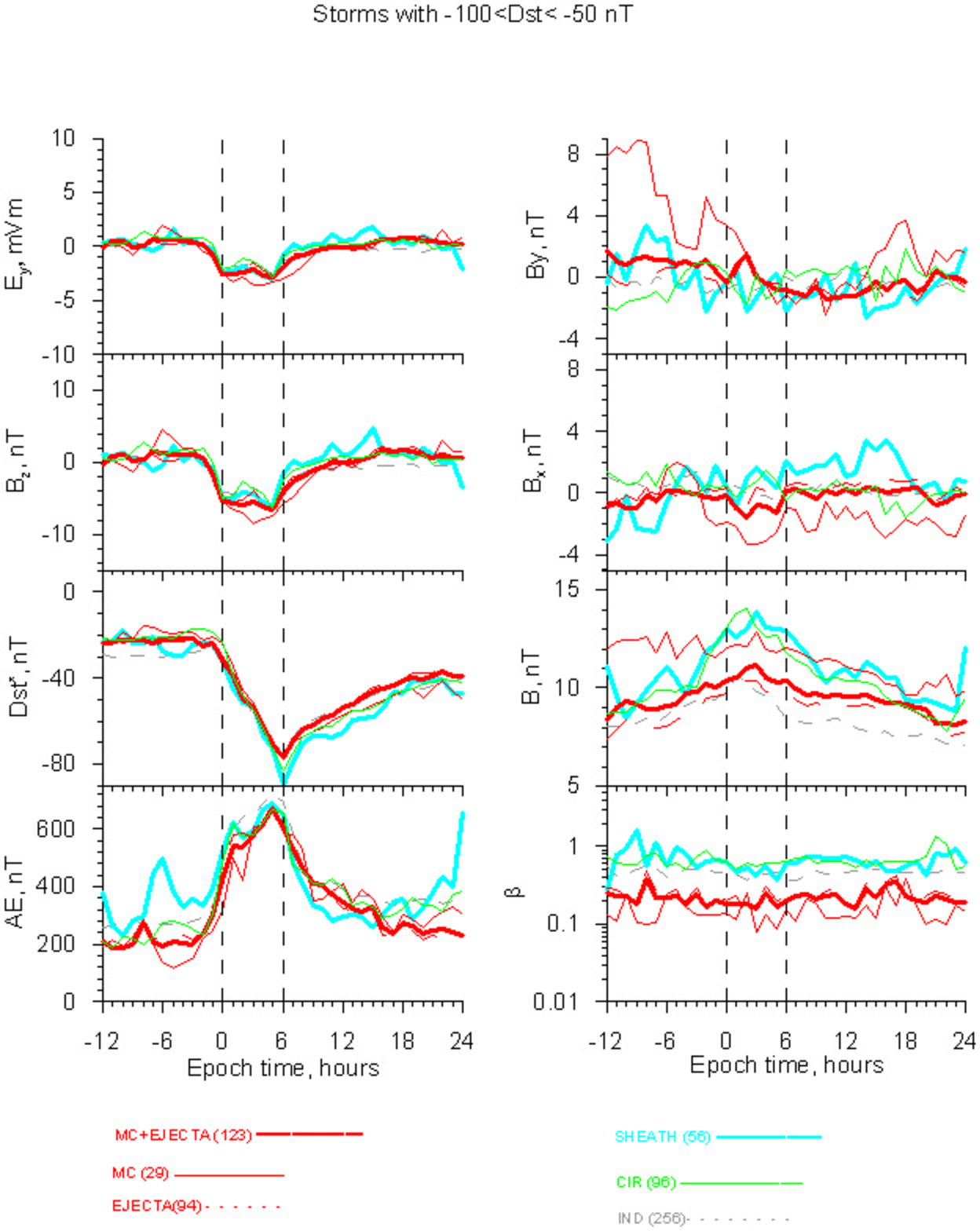}
\caption{The same as in Fig.2 for moderate storms with -100 $< Dst <$ -50 nT.}
\end{figure}

\begin{figure}
\includegraphics[width=0.8\textwidth]{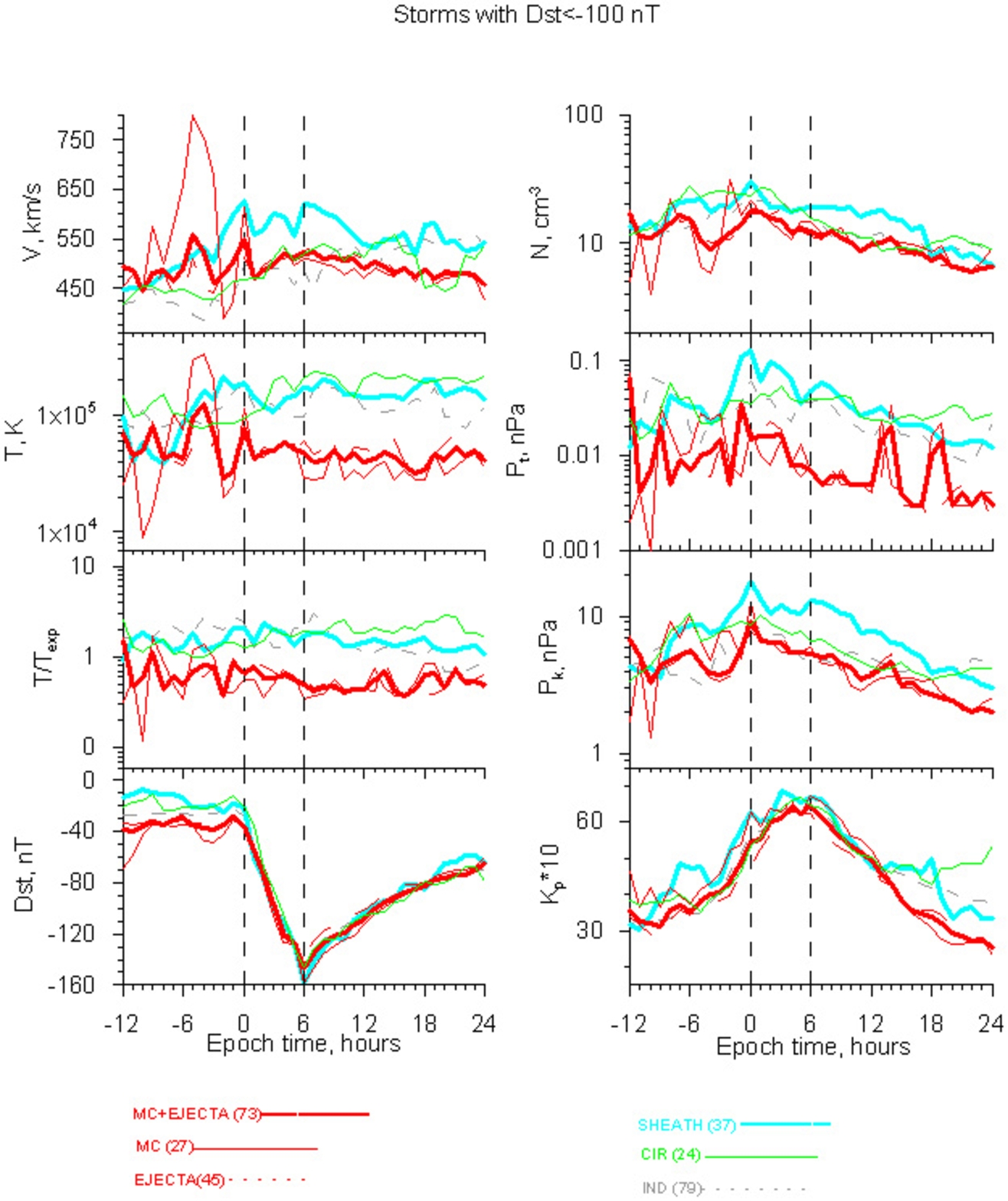}
\caption{The same as in Fig.1 for strong storms with $ Dst <$ -100 nT.}
\end{figure}

\begin{figure}
\includegraphics[width=0.8\textwidth]{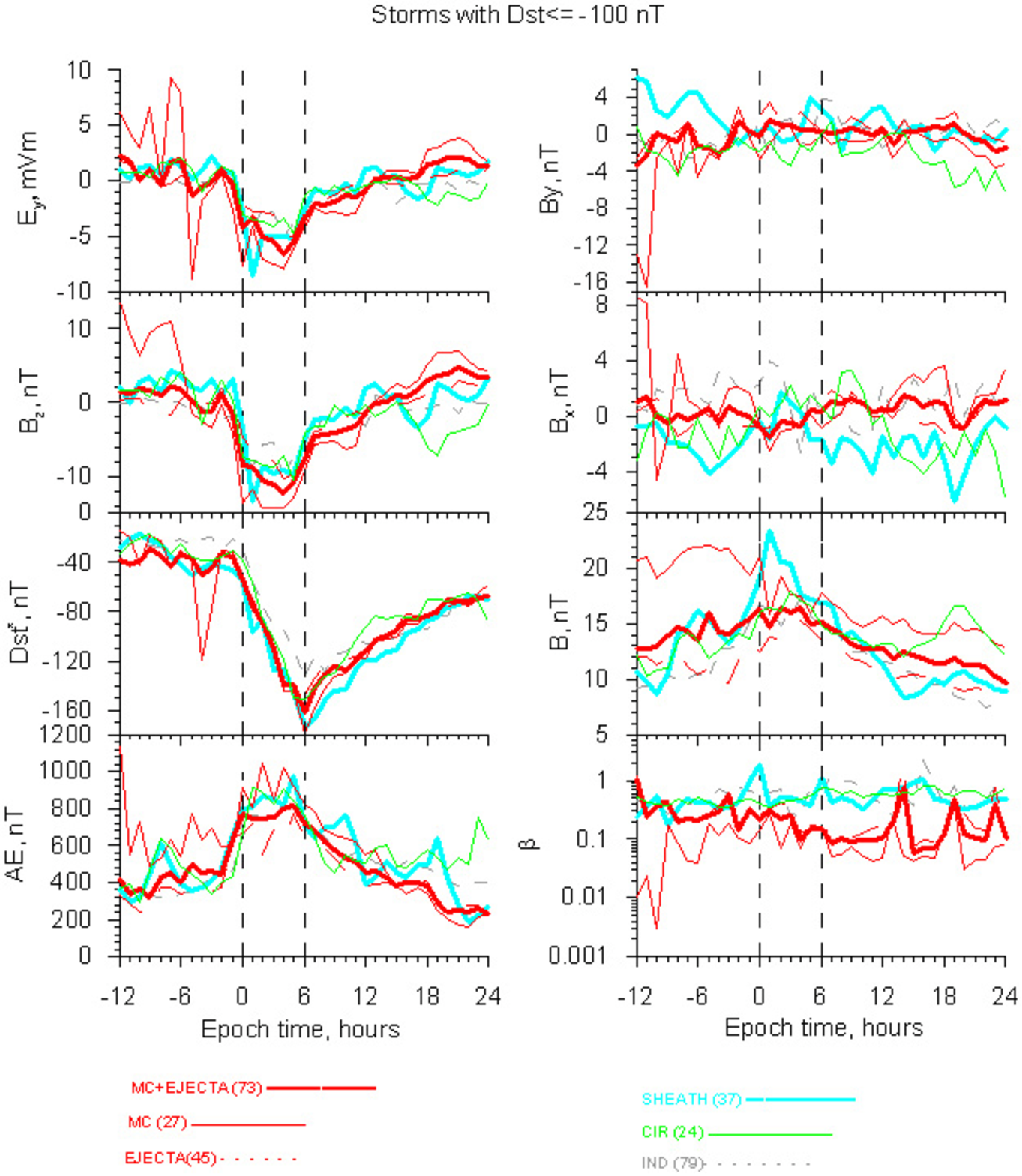}
\caption{The same as in Fig.2 for strong storms with $ Dst <$ -100 nT.}
\end{figure}

\begin{figure}
\includegraphics[width=0.8\textwidth]{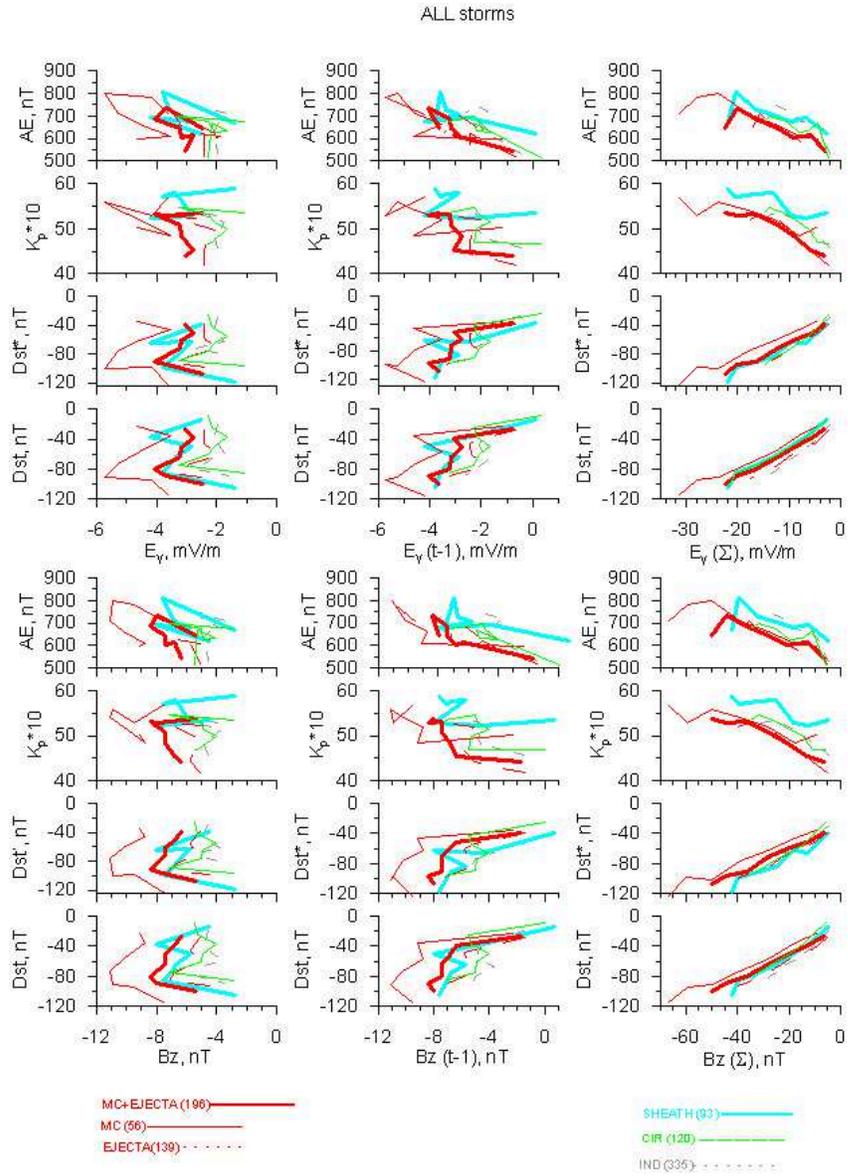}
\caption{Time evolution of $Dst, Dst^*, Kp$ and $AE$ indices during time evolution of $Bz$ and $Ey$ for time interval of "0" - "6" using 3 procedures (see text): 
(1. Left column) simultaneous measurements,
(2. Second column)  1-h displaced measurements,  and 
(3. Right column)  dependence of indices on integral value of sources.}
\end{figure}

\begin{figure}
\includegraphics[width=0.8\textwidth]{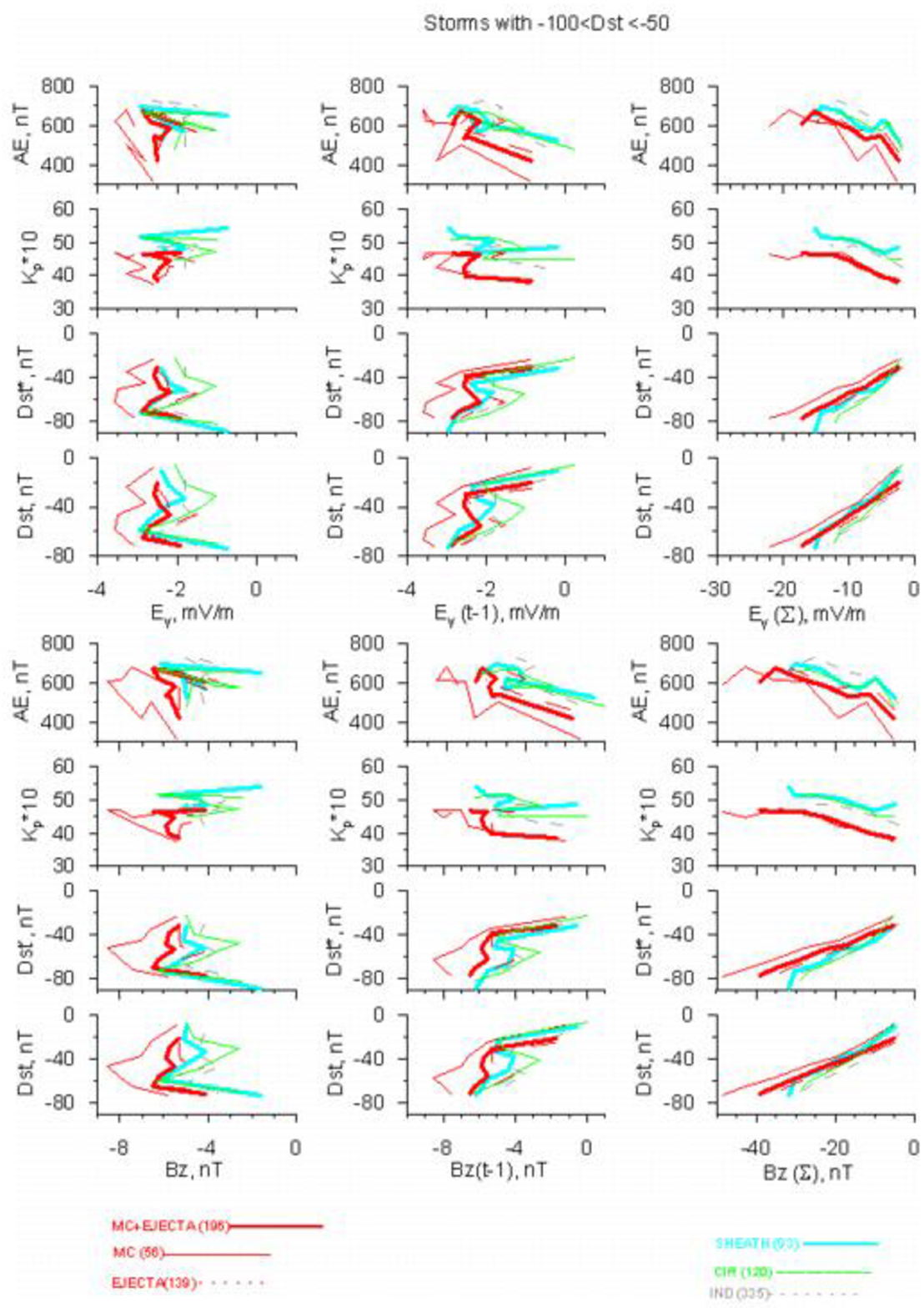}
\caption{The same as in Fig.7 for moderate storms with -100 $< Dst <$ -50 nT. }
\end{figure}

\begin{figure}
\includegraphics[width=0.8\textwidth]{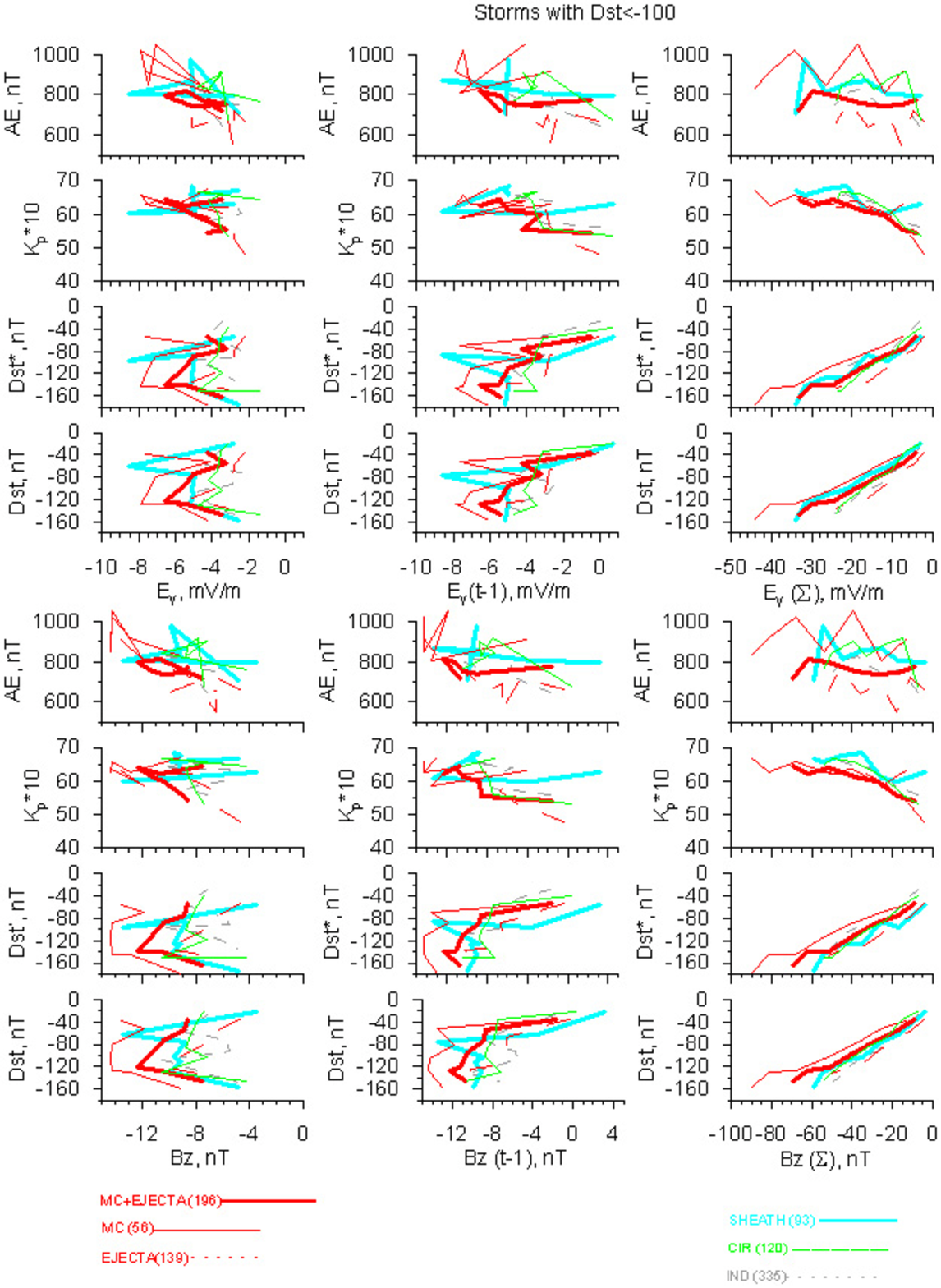}
\caption{The same as in Fig.7 for strong storms with $ Dst <$ -100 nT.}
\end{figure}

\end{document}